\documentclass[english]{article}
\usepackage{lmodern}
\usepackage{booktabs}
\pdfoutput=1

\usepackage[T1]{fontenc}
\usepackage[utf8]{inputenc}
\makeatletter

\usepackage{mathrsfs}   %\mathscr{L}
\usepackage{slashed}     %\slashed{p}
\usepackage{bbold}  % \mathbb{1} for the identity matrix
\usepackage{url}
\usepackage{graphicx}
\usepackage[colorlinks=true,linkcolor=redLinks,citecolor=greenLinks,urlcolor=redLinks, pdfborder={0 0 1}]{hyperref}
\usepackage{color} 
\usepackage[numbers,sort&compress]{natbib}
\usepackage{float}
\usepackage{multirow}
\usepackage{rotating}
\usepackage{amsmath}
\usepackage{amssymb}
\usepackage{breqn}
\usepackage{subcaption}
\usepackage{caption}
\usepackage{mciteplus}
\usepackage{authblk}
\usepackage{tikz}
\usepackage[normalem]{ulem}

\definecolor{greenLinks}{rgb}{0, 0.6, 0} 
\definecolor{blueLinks}{rgb}{0, 0, 0.6}
\definecolor{redLinks}{rgb}{0.6, 0, 0}
\definecolor{eprintLinks}{rgb}{0.4, 0.4, 0.4}
\definecolor{journalLinks}{rgb}{0.6, 0, 0}

\usepackage{multirow}
\let\orig@Hy@EveryPageAnchor\Hy@EveryPageAnchor
\def\Hy@EveryPageAnchor{%
    \begingroup
    \hypersetup{pdfview=Fit}%
    \orig@Hy@EveryPageAnchor
    \endgroup
}

\let\oldFootnote\footnote
\newcommand\nextToken\relax

\renewcommand\footnote[1]{%
    \oldFootnote{#1}\futurelet\nextToken\isFootnote}

\newcommand\isFootnote{%
    \ifx\footnote\nextToken\textsuperscript{,}\fi}

\makeatother

\newcommand{\fig}[1]{Fig.~\ref{#1}}

\newcommand{\sect}[1]{Sec.~\ref{#1}}

\usepackage{babel}
\begin{document}

\title{{\Large{}\vspace{-1.0cm}} \hfill {\normalsize{}IFIC/19-52}\\*[10mm]
  {\huge{}Sequentially loop suppressed fermion masses from a single discrete symmetry}{\Large{}\vspace{0.5cm}}}
\date{}

\author[1]{{\Large{}Carolina Arbel\'aez}\thanks{E-mail:  carolina.arbelaez@usm.cl}}
\author[1]{{\Large{}A. E. C\'arcamo Hern\'andez}\thanks{E-mail:  antonio.carcamo@usm.cl}}
\author[2]{{\Large{}Ricardo Cepedello}\thanks{E-mail:  ricepe@ific.uv.es}}
\author[3]{{\Large{}Sergey Kovalenko}\thanks{E-mail:  sergey.kovalenko@unab.cl}}
\author[1]{{\Large{}Ivan Schmidt}\thanks{E-mail:  ivan.schmidt@usm.cl}}

%%%%%%%%%%%%%%%%%%%%%%%%%%%%%%%%%%%%%%%%%%%%%%%%%%%%%%%%%%%%%%%%%%%%%%%%%%%%%%%%%%%%%%%%%%%%%%%%%%%%%%
% AFFILATIONS
\affil[1]{\small Universidad T\'ecnica Federico Santa Mar\'ia and Centro Cient\'ifico-Tecnol\'ogico de Valpara\'iso \protect\\ 
    Casilla 110-V, Valpara\'iso, Chile}
\affil[2]{\small AHEP Group, Instituto de F\'isica Corpuscular, CSIC - Universitat de Val\`encia \protect\\
    Edificio de Institutos de Paterna, Apartado 22085, E--46071 Val\`encia, Spain}
\affil[3]{\small Departamento de Ciencias F\'isicas, Universidad Andres Bello,
Sazi\'e 2212, Piso 7, Santiago, Chile}

\maketitle

\begin{center}
\today
\par\end{center}
\begin{abstract}
We propose a systematic and renormalizable sequential loop suppression mechanism to generate the hierarchy of the Standard Model fermion masses from one discrete symmetry. The discrete symmetry is sequentially softly broken in order to generate one-loop level masses for the bottom, charm, tau and muon leptons and two-loop level masses for the lightest Standard Model charged fermions. The tiny masses for the light active neutrinos are produced from radiative type-I seesaw mechanism, where the Dirac mass terms are effectively generated at two-loop level.  
\end{abstract}
\newpage{}

% Body of the text

\section{Introduction} 
\label{sec:intro}
Despite its striking consistency with experimental data, the Standard Model (SM) does not explain several fundamental issues, such as the number of fermion generations, the observed pattern of fermion masses and mixings, etc. In fact, the huge fermion mass hierarchy spans over a range of 13 orders of magnitude from the light active neutrino mass scale up to the top quark mass.

In the SM fermion masses arise from a Dirac mass term, which is completely invariant under the SM gauge group. Take for instance the Dirac term for the u-type quarks,
\begin{equation} \label{eq:dirac term u}
    \bar q_{L\alpha} \tilde{H} u_{R\beta}.
\end{equation}
Here $\tilde{H} = i \sigma^2 H^*$, $q_{L\alpha}$ is the left-handed quark doublet and $u_{R\beta}$ the right-handed u-quark singlet. As the three quark generations ($\alpha,\beta=1,2,3$) transform equally under the SM, all three masses will be generated by the same tree level Yukawa matrix after electroweak symmetry breaking (EWSB) and they should naturally have a similar value. However, in actual practice this is not the case as quark and lepton masses span several orders of magnitude. Thus, in order to accommodate the experimental values of the charged fermion masses in the SM, one needs to rely on an unnatural tuning among their different Yukawa couplings.

All these unexplained issues strongly indicate that new physics has to be invoked to address the fermion puzzle of the SM. To tackle the limitations of the SM, various extensions, including enlarged scalar and/or fermion sector, as well as extended symmetries, discrete or continuous, have been proposed in the literature \cite{Cheng:1977ir, Cheng:1980qt, Chang:1986bp, Mohapatra:1987nx, Branco:1987yg, Balakrishna:1988ks, Ma:1988fp, Ma:1989ys, Ma:1990ce, Ma:1998dn, Altmannshofer:2014qha, Ibarra:2014fla, Hernandez:2015hrt, Sierra:2016qfa, Arbelaez:2016mhg, Dev:2018pjn, Cepedello:2018rfh, CentellesChulia:2019gic, Bonilla:2018ynb, Avila:2019hhv, Arbelaez:2019wyz, CarcamoHernandez:2019lhv}. For instance, there have been attempts to explain the hierarchy of the SM via loops in well-known setups as the two Higgs doublet model \cite{Ibarra:2014fla} and supersymmetry \cite{Altmannshofer:2014qha}. Nevertheless, neither generate all the SM fermion masses of the quark and leptonic sector, including neutrinos.

A possible scenario to address the SM flavor puzzle is to effectively generate the masses of the different fermion generations via other mechanisms, such as radiative mass generation, in which different fermions have distinct loop order suppression. It is natural to start with the top mass generated at tree level, then the charm and bottom masses should appear at one-loop level, while the up, down and strange masses arise at two-loop level. For the leptonic sector, the muon and tau masses are generated at one-loop level, the electron mass at two-loop level, while the light active neutrinos have to acquire mass, at least, at four-loop level if they are of Dirac type, although this can be relaxed considering Majorana neutrinos with a seesaw-like mechanism \cite{Arbelaez:2019wyz}. For this realization to be possible, one needs an extra symmetry beyond the SM to differentiate between generations. This mechanism was proposed for the first time in \cite{CarcamoHernandez:2016pdu} in a non-renormalizable approach, while renormalizable versions have been implemented in \cite{CarcamoHernandez:2017cwi, CarcamoHernandez:2019cbd}. However, such renormalizable models rely on rather large scalar and fermion sectors,  sophisticated symmetries and lack a guiding principle for the model construction.\\

In this work we propose a systematic and renormalizable sequential loop suppression mechanism to generate the hierarchy of fermion masses from a single softly broken discrete $Z_n$ symmetry. The symmetry and the charges of the fields under it are chosen in such a way that each loop level breaks the discrete symmetry, but keeps a remnant exact subgroup of it, which is responsible for protecting the hierarchy.\\

The paper is organized as follows. In section \ref{sec:model} we describe how to generate a sequential loop suppression, using just one symmetry beyond the Standard Model. This mechanism is then implemented for the quark and leptonic sector using a $Z_4$ symmetry in sections \ref{sec:quarks} and \ref{sec:leptons}, respectively. In section \ref{sec:particles} we describe a set of possible charge assignments for the fields that mediate the quark and lepton radiative masses given in the previous sections. Furthermore, we shortly discuss the implications of the soft breaking mass terms of such models in section \ref{sec:softbreaking} and present our conclusions in section \ref{sec:discussion}. Finally, a different realization of the sequential loop suppression mechanism without the inclusion of the $U(1)_X$ symmetry is presented in  Appendix \ref{sec:app}.

\section{Loop suppression mechanism}\label{sec:model}

In this section we describe the general guidelines followed in order to generate a sequential loop suppression, using just one symmetry beyond the Standard Model. This mechanism is to be applied to the dynamical generation of Standard Model fermion masses, mimicking their hierarchy via loop orders. However, this framework goes beyond the application shown here regarding only the SM fermions and can be implemented in other scenarios.

For simplicity, we will use the operator $\mathcal{O}_\beta \equiv \bar{q}_{L\beta} \tilde{H} u_{R\beta}$ in \eqref{eq:dirac term u} to illustrate how the mechanism works.  The generalization for any non-diagonal Dirac flavor mass operator is straightforward. In the SM the operators $\mathcal{O}_1$, $\mathcal{O}_2$ and $\mathcal{O}_3$ are indistinguishable from a symmetry point of view, and for this reason, a new symmetry is needed to differentiate them. 
Then this symmetry has to be broken at some scale, allowing for the existence of the operators $\mathcal{O}_i$ in the low-scale theory, with suppression factors for each of them. The main strategies followed in the literature consist in working with the low energy scale directly, breaking softly the symmetries \cite{CarcamoHernandez:2016pdu}; or to generate these operators effectively by adding scalars that break spontaneously the new symmetries \cite{CarcamoHernandez:2017cwi, CarcamoHernandez:2019cbd}.

Here we restrict ourselves to the case of {\bf (i)} renormalizable Dirac Yukawa mass operators like \eqref{eq:dirac term u} and {\bf (ii)}  a new softly broken symmetry. For simplicity, we consider only abelian discrete symmetries. This new symmetry should forbid a Dirac Yukawa mass term at all orders, but once it is broken softly, an effective realization of this term via loops is allowed, while it is still forbidden at tree level, since it is a hard dim=4 term.  This same symmetry can also be used to distinguish between fermion generations, allowing the Dirac Yukawa mass operators to be realized at different loop orders for each generation. Nevertheless, a general implementation of this idea is sometimes difficult, if we consider just a single symmetry beyond the SM, since naive control of its breaking can spoil the desired features of the constructed model. For instance, when breaking softly the symmetry in order to allow $\mathcal{O}_2$, one must ensure that the same mass mechanism is not allowed for $\mathcal{O}_1$. Otherwise, it would not be possible to introduce the desired hierarchy to the quark mass spectrum. 

In this work we show a mechanism to generate effectively a mass hierarchy of operators among fermion generations, by softly breaking just one symmetry group $G$, while still protecting the hierarchy. Considering abelian discrete groups, the main requirement is that this extra symmetry group $G$ is not simple, and it contains at least one non-trivial normal subgroup $H$. Now consider that the two operators $\mathcal{O}_1$ and $\mathcal{O}_2$ transform non-trivially under the new symmetry $G$ as,
\begin{eqnarray} %\label{eq:}
    \mathcal{O}_1 &\sim& g \in G-H,\\
    \mathcal{O}_2 &\sim& h \in H-\{1\},
\end{eqnarray}
where $\{1\}$ is the identity group and $h \in H-\{1\}$ denotes any element of $H$ except the identity. Analogously, $g \in G-H$ refers to any element in $G$ but not in $H$.

In order to realize $\mathcal{O}_2$ at loop level, a soft breaking term transforming as $h^{-1} \in H-\{1\}$ is included. Because $H$ is a subgroup, no product of elements in $H$ can generate an element in $G-H$. For this reason, no combination of $\mathcal{O}_2$ with the soft breaking term can generate $\mathcal{O}_1$ at any level. To generate $\mathcal{O}_1$, a soft term transforming as $g^{-1} \in G-H$ is required. Consequently, the extra symmetry ensures that the operators $\mathcal{O}_2$ and $\mathcal{O}_1$ can be generated through different realizations even after the symmetry has been broken.\footnote{In principle a combination of the soft term transforming as $g^{-1}$ can generate a term transforming as $h^{-1} \in H-\{1\}$ and induce $\mathcal{O}_2$. However, this would be a higher order contribution because at least two soft-breaking terms will be needed instead of one, i.e. a subdominant contribution.} For instance, providing non-reducible diagrams \cite{Yao:2017vtm,CentellesChulia:2019xky}, $\mathcal{O}_2$ can be realized at one-loop level, while $\mathcal{O}_1$ can be constructed at a two-loop level.

It is worth mentioning the fact that after breaking the subgroup $H$ with the soft-breaking term the operator $\mathcal{O}_1$ cannot be realized, which means that there is a remnant symmetry protecting $\mathcal{O}_1$. This residual symmetry is the quotient group $G/H$.

Note that this reasoning can be applied to as many operators as long as the subsequent subgroups are not simple. For instance, following this idea one can systematically protect the loop realization of $n$ Dirac mass operators $\{ \mathcal{O}_1, \mathcal{O}_2, ..., \mathcal{O}_{n} \}$ if one finds a group $G$ such that contains the chain of normal subgroups $G \supset H_1 \supset H_2 \supset ... \supset H_{n-1}$ and forces the operators $\mathcal{O}_i$ to transform as elements of $H_{i-1} - H_{i}$ with $G \equiv H_0$ and $\{1\} \equiv H_{n}$. This will reproduce the hierarchy $\mathcal{O}_{n} > ... > \mathcal{O}_{1}$ by sequentially breaking softly the whole group $G$. The loop realization of $\mathcal{O}_{n}$ breaks only the subgroup $H_{n-1}$, $\mathcal{O}_{n-1}$ breaks $H_{n-2}$ and successively, until the $G$ is completely broken by the loop realization of $\mathcal{O}_{1}$.

Moving to a concrete example to illustrate this mechanism, the simplest groups for which a systematic chain of normal subgroups can be found are the cyclic $Z_n$ groups, with $n$ not a prime number.  Among them, the easiest of these chains is $Z_2 \subset Z_4 \subset Z_8 \subset ... \subset Z_{2^n}$. Other possibilities would be, for instance, $Z_3 \subset Z_9 \subset ... \subset Z_{3^n}$ or $Z_2 \subset Z_6 \subset Z_{12} \subset ...$. As well as for non-abelian groups as the symmetric group $S_n$, which always contains $S_{n-1}$ as a subgroup (for $n > 1$). As we want to apply this to the SM, we only need to generate masses up to the two-loop level, so we can take a $Z_4$ symmetry for simplicity. Considering as an example the u-mass operators $\mathcal{O}_\beta$ \eqref{eq:dirac term u}, we want to ensure that the mass of the top quark be generated at tree level, while the charm and up quarks arise at one- and two-loops, respectively. Following the arguments given above, for $G\equiv Z_4$ and $H\equiv Z_2$, so that  $Z_{4}/Z_{2}\sim Z'_{2}$, we have: 
\begin{itemize}
    \item $\mathcal{O}_3 \equiv \bar q_{L\alpha} \tilde{H} u_{R3}$ should transform as a singlet under $Z_4$, so the top mass can be realized at tree level. 
    \item On the other hand, $\mathcal{O}_2$ transform as $-1 \in Z_2-\{1\} \subset Z_4$ and $\mathcal{O}_1$ as $\pm i \in Z_4 - Z_2 \subset Z_4$. At the moment, neither of these two operators are allowed by the symmetries.
    \item A soft-breaking term, transforming as $-1$ under $Z_4$, is added in order to close the mass term for the charm quark $\mathcal{O}_2 \equiv \bar q_{L\alpha} \tilde{H} u_{R2}$ at one-loop level.
    \item This soft breaking term breaks $Z_4$ down to  a remnant $Z'_2$ group. The operator  $\mathcal{O}_1$ still cannot be realized since it transforms as $-1$ under $Z'_{2}$.
    \item A soft-breaking term, transforming as $-1$ under the remnant $Z'_2$ (or equivalently as $\mp i$ under the original $Z_4$), is then added to realize $\mathcal{O}_1 \equiv \bar q_{L\alpha} \tilde{H} u_{R1}$, breaking completely the initial $Z_4$ symmetry and generating the mass of the u-quark at two-loop level.
\end{itemize}
As a result, starting from a simple $Z_4$ symmetry, one can ensure that the mass mechanisms for the three u-quark families are decoupled, realized at different orders and protected through the sequential breaking of the cyclic $Z_4$ symmetry.

\section{Quark sector} \label{sec:quarks}

The idea that discrete symmetries are behind the pattern of the hierarchy of quark masses via loop suppression is not new in the literature \cite{CarcamoHernandez:2016pdu}. However, despite various interesting developments, the previous models require \emph{many} symmetries and assumptions in order to fix the specific fermionic patterns. In the next sections we illustrate the loop suppression mechanism explained before to generate the SM fermion mass hierarchy, based on a \emph{single} $Z_{4}$ symmetry. For the quark sector, we consider that the charm and bottom quark masses arise at one-loop level, while the up, down and strange quark masses are generated at two-loop level. On the other side, the top quark mass is generated at tree level from the standard renormalizable Yukawa operator.

\begin{figure}[h!]
    \centering
    \includegraphics[scale=0.9]{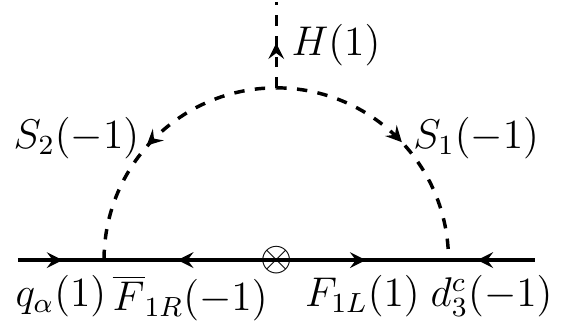}
        \hspace*{1cm}
    \includegraphics[scale=0.9]{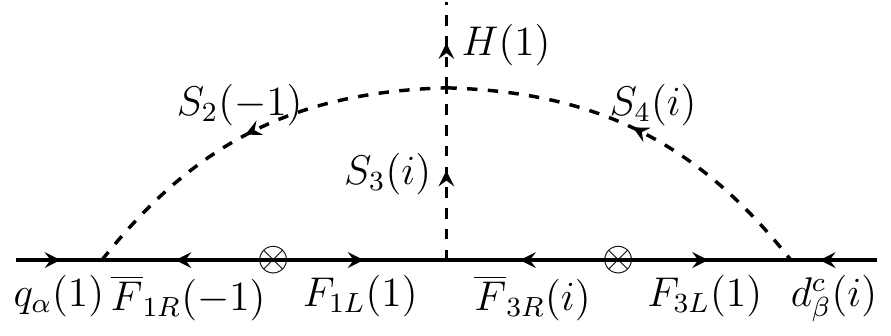}
    \caption{One- and two-loop Feynman diagrams contributing to the entries of the SM down-type quark mass matrix. Here $\alpha=1,2,3$ and $\beta=1,2$. The charge assignment under $Z_4$ is explicitly given in parentheses. This symmetry is broken softly by the vector-like masses denoted by $\otimes$.}
    \label{fig:quarksD}
\end{figure}

\begin{figure}[h!]
    \centering  
    \includegraphics[scale=0.9]{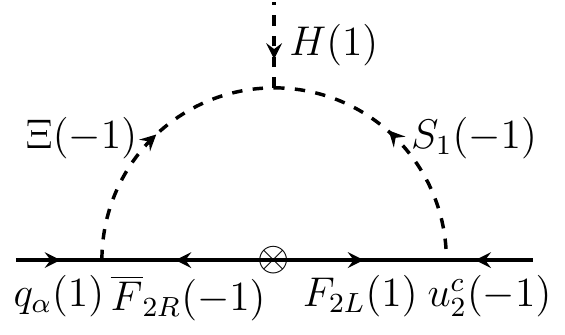}
        \hspace*{1cm}
    \includegraphics[scale=0.9]{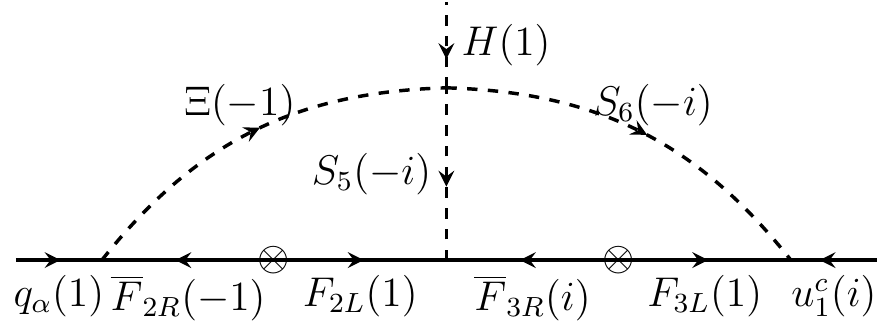}
\caption{One- and two-loop diagrams responsible for the masses of the SM up-type quark mass matrix. Here $\alpha=1,2,3$. The diagrams are generated once the discrete $Z_4$ symmetry is softly broken by the fermion masses, denoted by $\otimes$. The charges under $Z_4$ are given between parenthesis next to the corresponding field.}
    \label{fig:quarksU}
\end{figure}

The diagrams that generate the d-type and u-type quark masses are depicted in \fig{fig:quarksD} and \fig{fig:quarksU}, respectively. We give explicitly, between parenthesis, the charge of each field under $Z_4$, using as representation the 4-th roots of unity ($1$, $i$, $-1$, $-i$). The rest of the SM quantum numbers are shown in \sect{sec:particles}. It is straightforward to see that the vector-like masses ($\otimes$) of the new fermions $F$ break softly the $Z_4$ symmetry, allowing the loop to be closed. 

The role of the $Z_4$ symmetry and its subgroup $Z_2$ in the loop suppression mechanism can be directly seen from the diagrams. Take for instance \fig{fig:quarksU}. All the fields in the one-loop diagram, which generate the operator $\mathcal{O}_2 \equiv \bar{q}_{L\alpha} \tilde{H} u_{R2}$, transform as $\pm 1$ under $Z_4$, i.e. they all belong to the $Z_2$ subgroup. Meanwhile, the two-loop diagram ($\mathcal{O}_1$) contains new fields transforming as $\pm i$, which belong to the subset $Z_4-Z_2$, realizing the mechanism described in \sect{sec:model}.

The mass diagrams are generated by the following Yukawa interactions for the u-type and d-type quark sectors,
\begin{eqnarray}\label{LyU}
-\mathcal{L}_{Y}^{(u)} &=& \sum_{\alpha=1}^{3} Y_{\alpha}^{(u)} \, q_{\alpha} \, H \, u_{3}^{c}\notag\\
&& + \sum_{\alpha=1}^{3} y_{\alpha}^{(u)} \, q_{\alpha} \, \Xi^{\dagger} \, \overline{F}_{2R} 
+ x^{(u)} \, F_{2L} \, S_{1}^{\dagger} \, u_{2}^{c}  \notag \\
&& + \sum_{\gamma=1}^{2} \, w_{\gamma}^{(u)} \, F_{3L\gamma} \, S_{6} \, u_{1}^{c}
+ \sum_{\gamma=1}^{2} \, z_{\gamma}^{(u)} \, \overline{F}_{3R\gamma} \, S_{5} \, F_{2L} \; + \; H.c.,
\end{eqnarray}
\begin{eqnarray}\label{LyD}
-\mathcal{L}_{Y}^{(d)} &=& \sum_{\alpha=1}^{3} \sum_{\gamma=1}^{2} y_{\alpha\gamma}^{(d)} \, q_{\alpha} \, S_{2} \, \overline{F}_{1R\gamma}
+ \sum_{\gamma=1}^{2} x_{\gamma}^{(d)} \, F_{1L\gamma} \, S_{1} \, d_{3}^{c}  \notag \\
&& + \sum_{\beta,\gamma=1}^{2} w_{\gamma\beta}^{(d)} \, F_{3L\gamma} \, S_{4}^{\dagger} \, d_{\beta}^{c}
+ \sum_{\gamma,\delta=1}^{2} z_{\gamma\delta}^{(d)} \, \overline{F}_{3R\delta} \, S_{3}^{\dagger} \, F_{1L\gamma} \; + \; H.c.,
\end{eqnarray}
as well as the terms of scalar potential given by
\begin{equation} \label{eq:scalar quarks}
V \supset A^{(d)} \, H \, S_{1} \, S_{2} 
+ A^{(u)} \, H \, S_{1} \, \Xi
+ \kappa^{(d)} \, H^{\dagger} \, S_{2}^{\dagger} \, S_{3} \, S_{4} 
+ \kappa^{(u)} \, H \Xi \, S_{5}^{\dagger} \, S_{6}^{\dagger} \; + \; H.c.
\end{equation}
The number of generations considered for the new fermions is the minimal possible in order to reproduce all the SM quark masses.

Notice that the first line of \eqref{LyU} generates a tree level top quark mass. The terms on the second and third lines enters the one- and two-loop Feynman diagrams of \fig{fig:quarksU}, which generate the second and first columns of the u-type quark mass matrix, respectively, thus giving masses to the charm and up quarks. Similarly, for the SM d-type quark mass matrix, the terms on the first and second lines of \eqref{LyD} are associated to the one- and two-loop Feynman diagrams of \fig{fig:quarksD}, respectively, i.e. the different entries of the SM d-type quark mass matrix. This results in a one-loop level bottom quark mass and two-loop down and strange quark masses.

In order to close the loop Feynman diagrams, crucial for the implementation of sequential loop suppression mechanism, the following soft-breaking mass terms for the exotic fermions are required:
\begin{equation}
\mathcal{L}_{soft}^{(F)} = \sum_{\gamma ,\delta =1}^{2} M_{1\gamma\delta}^{(F)} \, \overline{F}_{1R\gamma} \, F_{1L\delta} 
+ M_{2}^{(F)} \, \overline{F}_{2R} \, F_{2L}
+ \sum_{\gamma,\delta=1}^{2} M_{3\gamma\delta}^{(F)} \, \overline{F}_{3R\gamma} \, F_{3L\delta} \; + \; H.c.
\end{equation}
The terms $\overline{F}_{1R}F_{1L}$ and $\overline{F}_{2R}F_{2L}$ break $Z_4$ in two units, leaving a $Z_2$ residual symmetry. These are associated to the one-loop diagrams. The two-loop diagrams break completely $Z_4$, as they are generated via the soft breaking term $\overline{F}_{3R}F_{3L}$, which transforms as $i$.

In principle, more soft breaking terms may arise depending on the SM quantum numbers of the fields for a specific model. Nevertheless, this is model dependent and cannot be treated in general. This does not represent any problem except for certain particular choices of the internal fields, for which an extra symmetry may be needed. This is the case for the particle content chosen in \sect{sec:particles}.
 
In what follows we will explain why the SM quark mass matrices given above are consistent with the quark masses and mixing angles. From the quark Yukawa interactions \eqref{LyU} and \eqref{LyD}, the SM u-type and d-type quark mass matrices are given by: 
\begin{equation}
M_{U} = \left( 
    \begin{array}{ccc}
        \widetilde{\varepsilon}_{11}^{(u)} & \varepsilon_{12}^{(u)} & Y_{1}^{(u)} \\
        \widetilde{\varepsilon}_{21}^{(u)} & \varepsilon_{22}^{(u)} & Y_{2}^{(u)} \\
        \widetilde{\varepsilon}_{31}^{(u)} & \varepsilon_{32}^{(u)} & Y_{3}^{(u)}
    \end{array}
\right) \frac{v}{\sqrt{2}},
\hspace{1cm}\hspace{1cm}
M_{D}=\left( 
    \begin{array}{ccc}
        \widetilde{\varepsilon}_{11}^{(d)} & \widetilde{\varepsilon}_{12}^{(d)} & \varepsilon_{13}^{(d)} \\
        \widetilde{\varepsilon}_{21}^{(d)} & \widetilde{\varepsilon}_{22}^{(d)} & \varepsilon_{23}^{(d)} \\
        \widetilde{\varepsilon}_{31}^{(d)} & \widetilde{\varepsilon}_{32}^{(d)} & \varepsilon_{33}^{(d)}
    \end{array}
\right) \frac{v}{\sqrt{2}},
\end{equation} 
where $\varepsilon$ and $\widetilde{\varepsilon}$ are dimensionless effective couplings generated at the one- and two-loop level, respectively, via the diagrams depicted in Figs.~\ref{fig:quarksD} and \ref{fig:quarksU}. The third column in $M_U$ is generated at tree-level from the first term in \eqref{LyU}.

In order to make explicit the loop nature of the hierarchy proposed here, we can parametrize the entries of both matrices as,
\begin{equation} \label{eq:paramet}
    Y_{i}^{(u)} \sim \mathcal{O}(1),
    \hspace{0.7cm}
    \widetilde{\varepsilon}_{i1}^{(u)} = \left( \frac{1}{16\pi^2} \right)^2 a_{i}^{(u)} \sim \lambda^8 \, \mathcal{O}(1),
    \hspace{0.7cm}
    \varepsilon_{i2}^{(u)} = \left( \frac{1}{16\pi^2} \right) b_{i}^{(u)} \sim \lambda^4 \, \mathcal{O}(1),
\end{equation}
\begin{equation*}
    \widetilde{\varepsilon}_{i1}^{(d)} = \left( \frac{1}{16\pi^2} \right)^2 a_{i}^{(d)} \sim \lambda^8 \, \mathcal{O}(1),
    \hspace{0.7cm}
    \widetilde{\varepsilon}_{i2}^{(d)} = \left( \frac{1}{16\pi^2} \right)^2 b_{i}^{(d)} \sim \lambda^6 \, \mathcal{O}(10),
    \hspace{0.7cm}
    \varepsilon_{i3}^{(d)} = \left( \frac{1}{16\pi^2} \right) c_{i}^{(d)} \sim 5 \lambda^3 \, \mathcal{O}(1),
\end{equation*}
Here $\lambda =0.225$ is one of the Wolfenstein parameters. For the model dimensionless parameters $a_{i}^{(u)}$, $b_{i}^{(u)}$, $Y_{i}^{(u)}$, $a_{i}^{(d)}$, $b_{i}^{(d)}$ and $c_{i}^{(d)}$ ($i=1,2,3$), which are complex in general, we indicated their order of magnitude, $\mathcal{O}(1)-\mathcal{O}(10)$, necessary to reproduce the observed values of the quark mixing angles. Therefore, a small hierarchy of one order of magnitude on the top of the loop suppression is still needed. However, the situation is considerably better compared to that of the SM, where significant paremeter tuning is required.

Concerning the SM quark mass spectrum and CKM parameters, these models have, in general, enough freedom to successfully accommodate the experimental values of the ten physical observables of the quark sector, i.e., the six quark masses, the three mixing angles and the Jarlskog invariant. Setting all the masses in the diagrams to order $10$ TeV, there are still $28$ dimensionless parameters between the elements of the Yukawa matrices and the scalar couplings, with values around $0.1$ and $1$. From \eqref{eq:paramet} we see that, apart from the hierarchy generated by the loop order, a small tuning in the parameters must be assumed in order to have a difference of one order of magnitude between the two-loop diagrams for the s- and d-quarks in Fig.~\ref{fig:quarksD} (right). This hierarchy can be easily accommodated, for instance, by imposing this same hierarchy in the corresponding entries of the Yukawas $y^{(d)}$, $w^{(d)}$ and $z^{(d)}$ in \eqref{LyD}.

\section{Leptonic sector} \label{sec:leptons}

Analogously to the quark sector, the same diagrams and $Z_4$ charges can be applied to generate the lepton masses. Of course, as the quantum number of the external particles are color blind, one should add either scalar leptoquarks or massive colorless vector-like fermions, as it will be discussed in \sect{sec:particles}.

\begin{figure}[h!]
    \centering
    \includegraphics[scale=0.9]{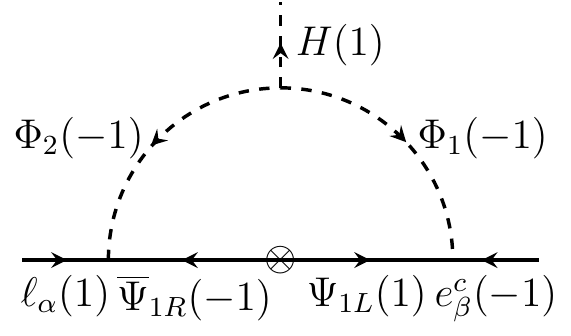}
        \hspace*{1cm}
    \includegraphics[scale=0.9]{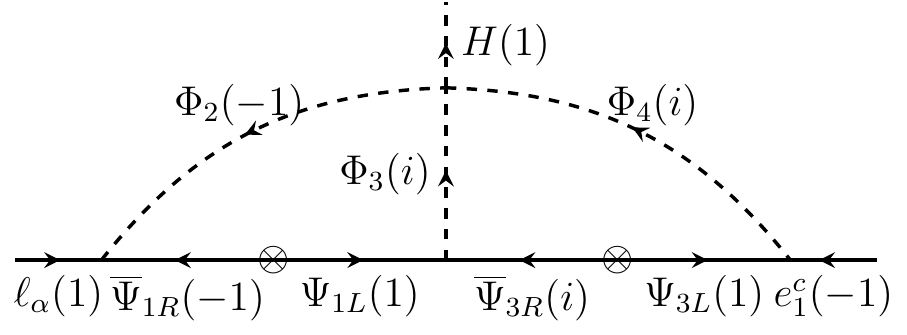}
    \caption{One- and two-loop Feynman diagrams contributing to the entries of the SM charged lepton mass matrix. Here, $\alpha=1,2,3$ and $\beta=2,3$. $Z_4$ charges are given explicitly in brackets and the softly breaking terms are denoted by $\otimes$.}
    \label{fig:chargedleptons}
\end{figure}

\begin{figure}[h!]
    \centering
    \includegraphics[scale=0.9]{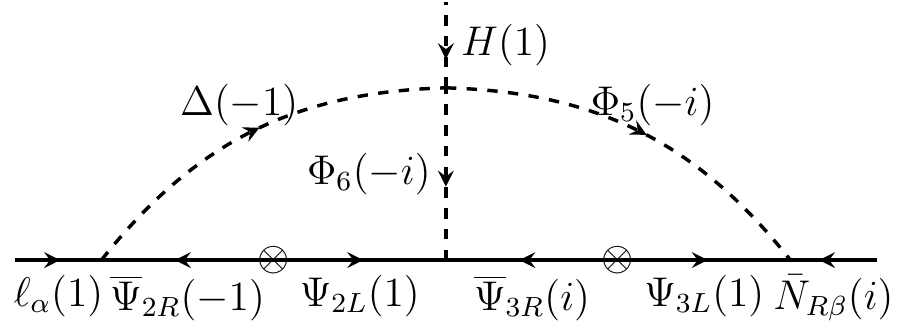}
    \caption{Two loop Feynman diagram that generates the Dirac neutrino mass term effectively. Here $\alpha=1,2,3$ and $\beta=1,2$. After softly breaking $Z_4$ the Majorana mass terms of \eqref{MNRmass} should be included, thus giving rise to an effective seesaw neutrino mass, see \fig{fig:eff seesaw}.}
    \label{fig:nudirac}
\end{figure}

The diagrams depicted in Figs.~\ref{fig:chargedleptons}~and~\ref{fig:nudirac} for the charged lepton and neutrino masses are generated by the following Yukawa interactions terms, respectively,
\begin{eqnarray}
-\mathcal{L}_{Y}^{(l)} &=& \sum_{\alpha=1}^{3} \sum_{\gamma=1}^{2} y_{\alpha\gamma}^{(l)} \, l_{\alpha} \, \Phi_{2} \, \overline{\Psi}_{1R\gamma}
+ \sum_{\gamma=1}^{2} \sum_{\beta=2}^{3} x_{\gamma\beta}^{(l)} \, \Psi_{1L\gamma} \, \Phi_{1} \, e_{\beta}^{c}  \notag \\
\label{eq:Nu-Yuk-1}
&& + \sum_{\gamma=1}^{2} \, w_{\gamma}^{(l)} \, \Psi_{3L\gamma} \, \Phi_{4}^{\dagger} \, e_{1}^{c}
+ \sum_{\gamma,\delta=1}^{2} \, z_{\gamma\delta}^{(l)} \, \overline{\Psi}_{3R\gamma} \, \Phi_{3}^{\dagger} \, \Psi_{1L\delta} \; +\; H.c.,
\end{eqnarray}
\begin{equation}
\label{eq:Nu-sector} 
-\mathcal{L}_{Y}^{\nu} = \sum_{\alpha=1}^{3} \sum_{\gamma=1}^{2} y_{\alpha\gamma}^{(\nu)} \, l_{\alpha} \, \Delta^{\dagger} \, \overline{\Psi}_{2R\gamma}
+ \sum_{\gamma,\delta=1}^{2} \, z_{\gamma\delta}^{(\nu)} \, \overline{\Psi}_{3R\gamma} \, \Phi_{6} \, \Psi_{2L\delta} 
+ \sum_{\gamma,\delta=1}^{2} \, w_{\gamma\delta}^{(\nu)} \, \Psi_{3L\gamma} \, \Phi_{5} \, N_{R\delta}^{c} \; + \; H.c.,
\end{equation}
as well as the relevant scalar interactions,
\begin{equation}
\label{eq:scalar-potential-1} 
V \supset A^{(l)} \, H^{\dagger} \, \Phi_{1}^{\dagger} \, \Phi_{2}^{\dagger}
+ \kappa^{(l)} \, H^{\dagger} \, \Phi_{2}^{\dagger} \, \Phi_{3} \, \Phi_{4}
+ \kappa^{(\nu)} \, H \, \Delta \, \Phi_{5}^{\dagger} \, \Phi_{6}^{\dagger} \; + \; H.c.
\end{equation}

Likewise the quark sector, here in order to close the loop diagrams we require the following soft-breaking mass terms 
\begin{equation}
\label{eq:soft-nu-sect-1}
\mathcal{L}_{soft}^{(\Psi)} = \sum_{\gamma,\delta=1}^{2} M_{1\gamma\delta}^{(\Psi)} \, \overline{\Psi}_{1R\gamma} \, \Psi _{1L\delta}
+ \sum_{\gamma,\delta=1}^{2} M_{2\gamma\delta}^{(\Psi)} \, \overline{\Psi}_{2R\gamma} \, \Psi_{2L\delta} 
+ \sum_{\gamma,\delta=1}^{2} \, M_{3\gamma\delta}^{(\Psi)} \, \overline{\Psi}_{3R\gamma} \, \Psi_{3L\delta} \; + \; H.c.,
\end{equation}
\begin{equation}\label{MNRmass}
\mathcal{L}_{soft}^{(N)} = \sum_{\gamma,\delta=1}^{2} M_{R\gamma\delta} \, \overline{N}_{R\gamma} \, N_{R\delta}^{c} \; + \; H.c.
\end{equation}
for the charged exotic fermions, $\Psi$, and right-handed Majorana neutrinos $N_{R}$. The number of generations of new fermions is the minimal phenomenologically allowed to reproduce the charged fermion masses and the neutrino masses. For the latter, only two neutrinos are massive, as the effective Dirac Yukawa coupling depicted in Fig.~\ref{fig:nudirac} is rank-2.

\begin{figure}[h!]
    \centering
    \includegraphics[scale=1]{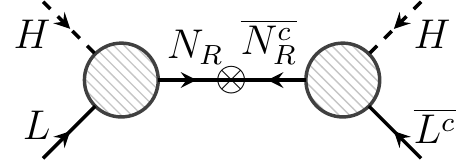}
    \caption{Radiative  seesaw Type-I mechanism for neutrinos \cite{Arbelaez:2019wyz}. The blob denotes the Dirac mass term generated at two-loop level through the diagram in \fig{fig:nudirac}. The symbol $\otimes$ indicates the Majorana mass for $N_R$ softly-breaking the $Z_4$. Generation indices are suppressed for simplicity.}
    \label{fig:eff seesaw}
\end{figure}

After the electroweak symmetry breaking the above-given Lagrangian (\ref{eq:Nu-Yuk-1})-(\ref{eq:scalar-potential-1}), along with the $Z_4$ soft-breaking mass terms (\ref{eq:soft-nu-sect-1}), (\ref{MNRmass}), lead to one-loop level masses for the tau and muon, and a two-loop mass for the electron, shown in \fig{fig:chargedleptons}. A two-loop Dirac mass term for neutrinos arises from the Feynman diagram in \fig{fig:nudirac}, analogously to the Dirac mass for the up quark, \fig{fig:quarksU}. In principle, for similar values of the couplings and masses, the neutrino mass would be roughly of the order of the electron mass, as both are generated through a similar mechanism. However, since the initial $Z_4$ symmetry is completely broken, a Majorana mass term is allowed for the fermions $N_{R}$ in \eqref{MNRmass}. Therefore, one can generate an effective four-loop Majorana mass for the light active neutrinos via the radiative seesaw Type-I  mechanism shown in \fig{fig:eff seesaw} and 
first described in \cite{Arbelaez:2019wyz}. Given the seesaw relation for the lightest neutrinos
\begin{equation}
    m_{\nu}\simeq m_{D}^{T} M_{R}^{-1} m_{D},
    \label{eq:numass}
\end{equation}
and since $m_{D}$ is generated at two-loop level by the diagram \fig{fig:nudirac}, the light active neutrino mass scale comes with a factor of $(\frac{1}{16\pi^2})^4$, which could guarantee a strong neutrino mass suppression even for a relatively small Majorana mass scale $M_R$. 

The Dirac mass can be computed straightforwardly from the diagram of \fig{fig:nudirac} and the neutrino Yukawa terms given in (\ref{eq:Nu-sector}), with the result:
\begin{equation}
(m_{D})_{\alpha \beta} = \left( \frac{1}{16\pi^2} \right)^2 
\kappa^{(\nu)}v \, \left[ y_{\alpha\gamma}^{(\nu)} z_{\delta\sigma}^{(\nu)} w_{\rho\beta}^{(\nu)} \right] \,  
\frac{M_{3\sigma\rho}^{(\Psi)} }{{M_{2\gamma\delta}^{(\Psi)}}} \; I_{loop}
\label{eq:mD1}
\end{equation}
where $v \equiv \langle H_{0} \rangle$ and $I_{loop}$ is a dimensionless two-loop function.

Here we are only interested in estimating the characteristic neutrino mass scale of this model. For this reason, we consider the unrealistic case with one massive neutrino, by assuming no hierarchy or flavor structure in the Yukawa couplings and just one copy of the new fermions, i.e.  $y_{\alpha\gamma}^{(\nu)} \equiv z_{\delta\sigma}^{(\nu)} \equiv w_{\rho\beta}^{(\nu)} \equiv Y$.

Moreover, as the masses of the $\Psi_2$, $\Psi_3$ and $N_R$ exotic fermionic fields are associated with the softly breaking scale of $Z_4$, we consider, for the sake of simplicity, a scenario where they are diagonal and degenerate, i.e., $M_{3\sigma\rho}^{(\Psi)}=M_{2\gamma\delta}^{(\Psi)}=M_R$, and also for the scalar fields mediating the radiative processes, $m_{\Phi_5}=m_{\Phi_6}=m_S$. Then, the Dirac neutrino mass scale given by \eqref{eq:mD1} can be estimated as,
\begin{equation}
m_{D} \sim \left( \frac{1}{16\pi^2} \right)^2 
\kappa^{(\nu)}
v Y^3 \; I_{loop}(m_S/M_R),
\end{equation}
where now the loop integral is a dimensionless function of just the ratio of the scalar/fermion masses given by
\begin{equation}
    I_{loop}(x) = (16\pi^2)^2 \int \frac{d^{4}k}{(2\pi)^4} \int \frac{d^{4}q}{(2\pi)^4} \frac{1}{ (k^2-1) (k^2-x^2) (q^2-1) (q^2-x^2) ((k+q)^2-x^2) }.
\end{equation}
This integral can be decomposed in terms of one master integral, for which an analytical solution can be easily found \cite{Sierra:2014rxa,Martin:2016bgz}.

\begin{figure}
    \centering
    \includegraphics[width=0.7\textwidth]{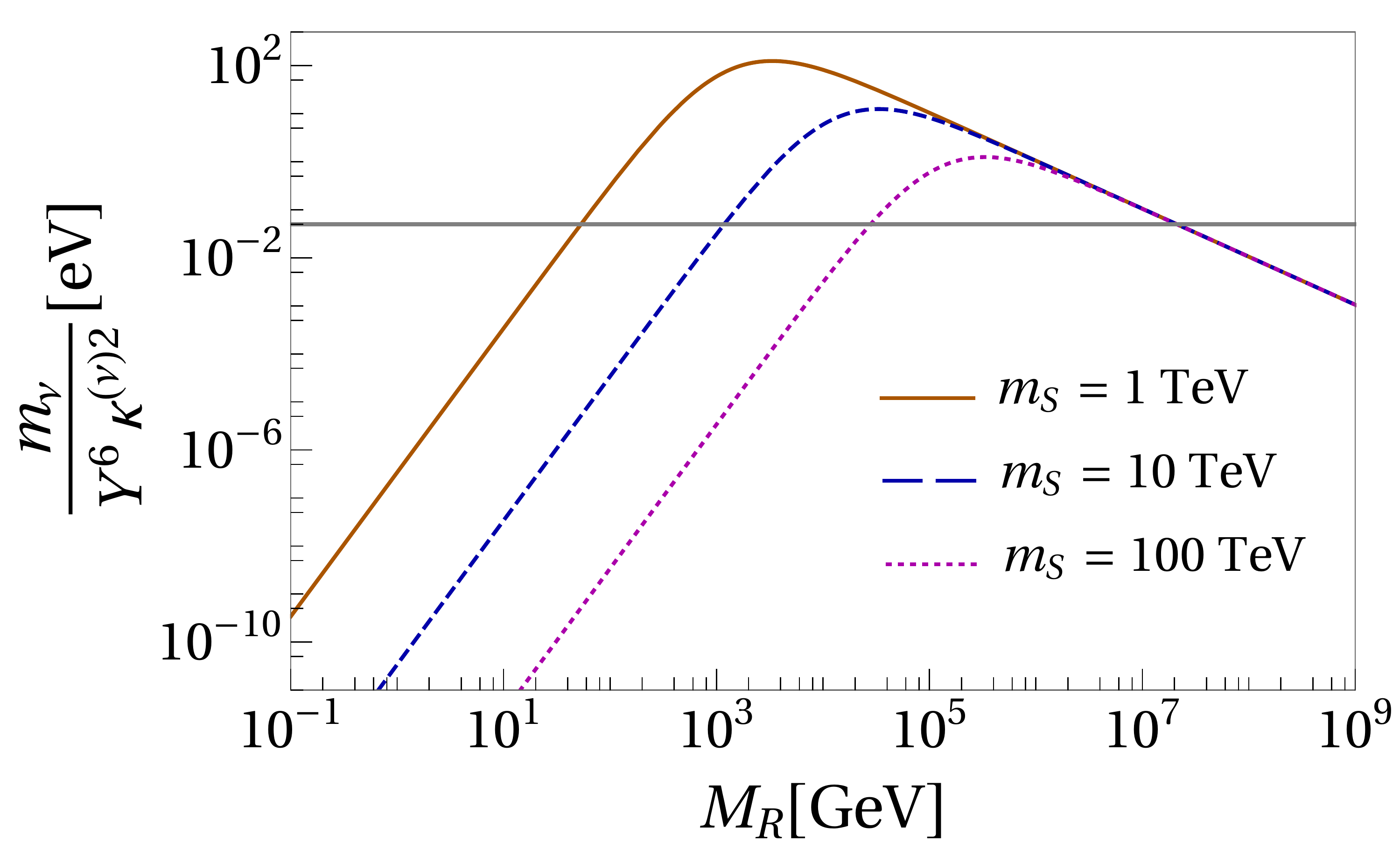} 
    \caption{Neutrino mass scale for different choices of the scalar masses. Here, we made simplifying assumptions $M_{2\gamma\delta}^{(\Psi)} = M_{3\sigma\rho}^{(\Psi)} = M_{R}$ for the exotic fermion masses and $m_{\Phi_{5}}=m_{\Phi_{6}}=m_{S}$ for the scalars. The grey horizontal line denotes the atmospheric mass scale, i.e: $m_{\nu}\sim 0.05$ eV.} 
    \label{fig:manu}
\end{figure}

The mass scale of the lightest active neutrino can be directly estimated through the seesaw relation shown in \eqref{eq:numass}, yielding the following estimate,
\begin{equation}
\label{eq:Seesaw-1}
m_{\nu} \sim \left ( \frac{1}{16 \pi ^2} \right )^4 
(\kappa^{(\nu)})^{2}
%\kappa_{4}^{2} 
\, Y^6 \frac{v^2}{M_R} \; \left[I_{loop}(m_S/M_R)\right]^2.
\end{equation}
This mass scale as a function of $M_R$  is plotted in \fig{fig:manu} for several scalar masses. We can observe that the difference, compared with the standard seesaw type-I/III, arises from the fact that the neutrino mass $m_{\nu}$ is considerably suppressed even for small values of the Majorana mass $M_{R}$, due to the effective four-loop level suppression. For example, if we set the values of the couplings to  $\mathcal{O}(1)$, one can fit the atmospheric neutrino oscillation scale $m_{\nu}\sim 0.05$ eV \cite{deSalas:2017kay} with the relatively small Majorana mass of the order of $100$~TeV. This should be compared to the typical Majorana mass scale of $10^{14}$ GeV in the standard seesaw mechanism with order 1 couplings.

It is worth mentioning that taking into account the limits from big bang nucleosynthesis \cite{Deppisch:2015qwa} and $\Delta N_{\text{eff}}$ \cite{Gariazzo:2019gyi}, which set a rough lower bound of $M_R \gtrsim (0.1 - 1)$ GeV, this class of models are severely constrained \cite{Arbelaez:2019wyz}. From \fig{fig:manu} we can see that even for the scalar masses of order $1$ TeV and couplings $\mathcal{O}(1)$, in order to get a reasonable neutrino mass without large non-perturbative values of the couplings in (\ref{eq:Seesaw-1}), the values of $M_R$ should lie above $10$ GeV. While an upper limit of $10^7$ GeV holds for scalars masses roughly below this mass.

Experimental bounds from the searches for lepton number violation are applicable for the range $100$~MeV $\leq M_{R} \leq$ 500 GeV \cite{Lindner:2016bgg}, depending on the value $m_S$. Also limits on the Yukawa couplings and $M_R$ are available from the experimental searches for Lepton flavor violating processes, like $\mu \rightarrow e \gamma$ \cite{Cai:2017jrq}. However, our goal here is to provide an order-of-magnitude estimate of the characteristic neutrino mass scale for this model without going into details. A more insightful study, including fitting neutrino oscillation data, should be done in order to set well-grounded limits to this type of neutrino mass models.

Analogous to the quark sector, our model can accommodate the experimental data of the SM lepton masses and mixings. As shown in \cite{Arbelaez:2019wyz}, for a new physics mass scale of order TeV, neutrino oscillation data can be directly fitted with $\mathcal{O}(1)$ couplings. 

Similarly, for the charged leptons, given their masses and their corresponding factors of $1/(16\pi^2)$ arising from the loop diagrams, we can roughly estimate that our model can reproduce the SM lepton hierarchy with couplings between $\mathcal{O}(0.1)$ and $\mathcal{O}(1)$, comparable to the quark sector.

\section{Particle content} \label{sec:particles}

In this section, we describe the charge assignments for all the fields mediating the previously discussed radiative processes, which give masses to the SM quarks lighter than the top quark and to the SM leptons. We discuss different scenarios, characterized by the specific nature of the internal fields mediating the loop. Table~\ref{tab:particles1} gives the SM and $N_{R}$ charge assignments under the $Z_{4}$ discrete group. These fields are common to all the models. In Tables \ref{tab:particles2}, \ref{tab:particles3} and \ref{tab:particles4}, the extra particle content used in each model on top of the SM fields is described. In \emph{Comments}, the possible nature of the fields inside the loop is given, in order to provide an insight to the phenomenology of these particles. Depending on the $SU(3)$ and $SU(2)$ charge assignments, the new fermions $F_{L/R}$ and $\Psi_{L/R}$ can be considered either as vector-like colored fields or vector-like non-colored fields. At the same time, the scalar sector could be colored or non-colored too. All the quantum numbers are common for all the generations of a given field, except for the $Z_4$ charges, which are generation dependent. \\

As described in the previous sections, in order to generate the operators, which give masses to the quark and leptons radiatively, we should softly break the discrete symmetry inside the loops, in our case $Z_{4}$ or the remnant $Z_{2}$. Once the symmetry of the model is broken, for consistency we should add to the theory all the possible soft breaking terms. Some of these terms, combined with the interactions of the SM fields, the extra fermions and the scalars inside the loop, may be dangerous in the sense that they could allow the SM quarks and leptons to acquire masses at tree level. Therefore, if we want the loop diagrams to be leading contributions to the fermion masses, we must forbid the tree level mass operators by an extra symmetry such as a global $U(1)_{X}$. Note that the necessity of this $U(1)_{X}$ is a model dependent feature due to our choice of the SM quantum numbers of the particles in the loop and it is not related to the above-described loop suppression mechanism based on a discrete symmetry, which is model independent.

\begin{table}[h!]
\begin{center}
\begin{tabular}{| c | c | c |}
  \hline
  \hspace{0.1cm}  Fields     \hspace{0.1cm}       &  \hspace{0.4cm}  $SU(3)_C \times SU(2)_L \times U(1)_Y$     \hspace{0.4cm}    & \hspace{0.4cm}   $Z_4$            \hspace{0.4cm} \\
\hline \hline
    $q_\alpha$   &       ($\mathbf{3}$, $\mathbf{2}$,  1/6) &  ($1$, $1$, $1$) \\
    $u^c_\alpha$ & ($\mathbf{\bar{3}}$, $\mathbf{1}$, -2/3) & ($i$, $-1$, $1$) \\
    $d^c_\alpha$ & ($\mathbf{\bar{3}}$, $\mathbf{1}$,  1/3) & ($i$, $i$, $-1$) \\
\hline \hline
    $\ell_\alpha$ & ($\mathbf{1}$, $\mathbf{2}$,  -1/2) &  ($1$, $1$, $1$) \\
    $e^c_\alpha$ & ($\mathbf{1}$, $\mathbf{1}$,    1) &  ($i$, $-1$, $-1$) \\
    $N_{R\beta}$ & ($\mathbf{1}$, $\mathbf{1}$,   0) &  ($i$, $i$) \\
\hline \hline
    $H$ & ($\mathbf{1}$, $\mathbf{2}$,  1/2) &  $1$ \\
    \hline
  \end{tabular}
\end{center}
\caption{Charge assignments for the SM fields and the right-handed field $N_R$, common to all the models. Not all the fields are charged under the extra global $U(1)_{X}$ symmetry. We consider the standard number of generations $\alpha=1,2,3$, while $\beta=1,2$ , to give masses to two active light neutrinos, which is the minimal phenomenologically viable choice. In the last column in the brackets we indicate $Z_{4}$ charges for the three generations.}
 \label{tab:particles1}
\end{table}%

\subsection{Option 1: No colored scalars} \label{sec:particles 1}

Let the new scalars be $SU(3)_C$ singlets, so the same set of scalars can, in principle, mediate both quark and lepton masses. In this case, specified in Table~\ref{tab:particles2}, the number of scalars is considerably reduced in comparison to the case displayed in Figs~\ref{fig:quarksD}~-~\ref{fig:nudirac}. Here, the scalars mediating the quark masses, $S_i$, can be identified with those participating in the generation of the lepton masses, i.e. $S_i \equiv \Phi_i$. Moreover, we may set $S_6 = S_5$ ($\Phi_6 = \Phi_5$) and $\Xi=\Delta=S_2$, further simplifying the scalar content. On the other side, this scenario requires vector-like colored and non-colored fermion fields, participating separately in the quark and lepton masses.

\begin{table}[h!]
\begin{center}
\begin{tabular}{| c || c | c | c | c || c |}
  \hline
  &  \hspace{0.1cm} Fields \hspace{0.1cm}  &  \hspace{0.4cm} $SU(3)_C \times SU(2)_L \times U(1)_Y$ \hspace{0.4cm}  &  \hspace{0.4cm} $Z_4$ \hspace{0.4cm}  &  \hspace{0.4cm} $U(1)_{X}$ \hspace{0.4cm}  &  \hspace{0.4cm} Comments \hspace{0.4cm}  \\
\hline \hline
\multirow{6}{*}{ \begin{turn}{90} Fermions \end{turn} }
    &    ($F_{1L}$,    $F_{1R}$) &  ($\mathbf{3}$, $\mathbf{1}$,  2/3) &  ($1$, $-1$) &     $x$ & Vector-like $u$-quark \\
    &    ($F_{2L}$,    $F_{2R}$) &  ($\mathbf{3}$, $\mathbf{1}$, -1/3) &  ($1$, $-1$) &    $-x$ & Vector-like $d$-quark \\
    &    ($F_{3L}$,    $F_{3R}$) &  ($\mathbf{3}$, $\mathbf{2}$,  1/6) &  ($1$, $-i$) &  $-x/2$ & Vector-like doublet $Q$ \\ 
    & ($\Psi_{1L}$, $\Psi_{1R}$) &  ($\mathbf{1}$, $\mathbf{1}$,    0) &  ($1$, $-1$) &     $x$ & Vector-like $\nu_R$ \\
    & ($\Psi_{2L}$, $\Psi_{2R}$) &  ($\mathbf{1}$, $\mathbf{1}$,    1) &  ($1$, $-1$) &    $-x$ & Vector-like $e_R$ \\
    & ($\Psi_{3L}$, $\Psi_{3R}$) &  ($\mathbf{1}$, $\mathbf{2}$, -1/2) &  ($1$, $-i$) &  $-x/2$ & Vector-like doublet $L$ \\
\hline \hline
\multirow{5}{*}{ \begin{turn}{90} Scalars \end{turn} } 
    & $S_1$ & ($\mathbf{1}$, $\mathbf{1}$,   -1) &  $-1$ &    $-x$ & Scalar charged singlet \\
    & $S_2$ & ($\mathbf{1}$, $\mathbf{2}$,  1/2) &  $-1$ &     $x$ & Inert doublet \\
    & $S_3$ & ($\mathbf{1}$, $\mathbf{2}$,  1/2) &   $i$ &  $3x/2$ & Inert doublet \\
    & $S_4$ & ($\mathbf{1}$, $\mathbf{2}$,  1/2) &   $i$ &  $-x/2$ & Inert doublet \\
    & $S_5$ & ($\mathbf{1}$, $\mathbf{2}$,  1/2) &  $-i$ &   $x/2$ & Inert doublet \\
    \hline
  \end{tabular}
\end{center}
\caption{Charge assignments for the fermions and scalars contributing to the diagrams Figs.~\ref{fig:quarksD}-\ref{fig:nudirac}, considering that all the scalars are color blind. In the brackets of the $Z_{4}$-column we indicate the corresponding charges of the left-handed and right-handed vector-like fermions, respectively. The $U(1)_{X}$ global symmetry and the field charges are discussed in section \ref{sec:softbreaking}.}
 \label{tab:particles2}
\end{table}%

It is worth mentioning that this scenario will have a rich phenomenology, containing vector-like copies of all the SM fermions, for which there is an extensive literature on experimental searches and phenomenological studies \cite{Falkowski:2013jya, Kumar:2015tna, Aaboud:2017zfn, Aaboud:2018saj, Aaboud:2018pii}

\subsection{Option 2: All vector-like fields are colored}

Here we consider the possibility that all the vector-like fermion fields carry a non-trivial $SU(3)$ color charge. In this case, the same specified in Table~\ref{tab:particles3} fermions that mediate the quark masses can also enter the lepton masses, i.e. $\Psi_i \equiv F_i$. In contrast to the previous example, here one needs to introduce color charged scalars in order to generate lepton masses. For this case also, $\Xi = S_2$.

\begin{table}[h!]
\begin{center}
\begin{tabular}{| c || c | c | c | c || c |}
  \hline
  &  \hspace{0.1cm} Fields \hspace{0.1cm}  &  \hspace{0.4cm} $SU(3)_C \times SU(2)_L \times U(1)_Y$ \hspace{0.4cm}  &  \hspace{0.4cm} $Z_4$ \hspace{0.4cm}  &  \hspace{0.4cm} $U(1)_{X}$ \hspace{0.4cm}  &  \hspace{0.4cm} Comments \hspace{0.4cm}  \\
\hline \hline
\multirow{3}{*}{ \begin{turn}{90} Fermions \end{turn} }
    &  ($F_{1L}$, $F_{1R}$)  &  ($\mathbf{3}$, $\mathbf{1}$,  2/3)  &  ($1$, $-1$)  &     $x$  & Vector-like $u$-quark \\
    &  ($F_{2L}$, $F_{2R}$)  &  ($\mathbf{3}$, $\mathbf{1}$, -1/3)  &  ($1$, $-1$)  &    $-x$  & Vector-like $d$-quark \\
    &  ($F_{3L}$, $F_{3R}$)  &  ($\mathbf{3}$, $\mathbf{2}$,  1/6)  &  ($1$, $-i$)  &  $-x/2$  & Vector-like doublet $Q$ \\ 
    &&&&&\\
\hline \hline
\multirow{12}{*}{ \begin{turn}{90} Scalars \end{turn} } 
    &  $S_1$ & ($\mathbf{1}$, $\mathbf{1}$,   -1) &  $-1$ &    $-x$ & Scalar charged singlet \\
    &  $S_2$ & ($\mathbf{1}$, $\mathbf{2}$,  1/2) &  $-1$ &     $x$ & Inert doublet \\
    &  $S_3$ & ($\mathbf{1}$, $\mathbf{2}$,  1/2) &   $i$ &  $3x/2$ & Inert doublet \\
    &  $S_4$ & ($\mathbf{1}$, $\mathbf{2}$,  1/2) &   $i$ &  $-x/2$ & Inert doublet \\
    &  $S_5$ & ($\mathbf{1}$, $\mathbf{2}$,  1/2) &  $-i$ &   $x/2$ & Inert doublet \\
    \cline{2-6}
    &&&&&\\[-3mm]
    &  $\Phi_1$ & ($\mathbf{\overline{3}}$, $\mathbf{1}$, -5/3) &  $-1$ &    $-x$ &  - \\
    &  $\Phi_2$ &            ($\mathbf{3}$, $\mathbf{2}$,  7/6) &  $-1$ &     $x$ &  - \\
    &  $\Phi_3$ &            ($\mathbf{1}$, $\mathbf{2}$,  1/2) &   $i$ &  $3x/2$ & Inert doublet \\
    &  $\Phi_4$ &            ($\mathbf{3}$, $\mathbf{2}$,  7/6) &   $i$ &  $-x/2$ &  - \\
    &  $\Phi_5$ & ($\mathbf{\overline{3}}$, $\mathbf{2}$, -1/6) &  $-i$ &   $x/2$ & Scalar $Q$ \\
    &  $\Phi_6$ &            ($\mathbf{1}$, $\mathbf{2}$,  1/2) &  $-i$ &   $x/2$ & Inert doublet \\
    &  $\Delta$ & ($\mathbf{\overline{3}}$, $\mathbf{2}$, -1/6) &  $-1$ &     $x$ & Scalar $Q$ \\
    \hline
  \end{tabular}
\end{center}
\caption{Charge assignments for the fermions and scalars that participate in the diagrams \mbox{Figs.~\ref{fig:quarksD}-\ref{fig:nudirac}} considering that the same set of colored fermions mediate quark and lepton masses, i.e. $\Psi_i \equiv F_i$. Note that in this case colored scalars with $U(1)_{X}$ charges appear. Other notations are the same as in Table~\ref{tab:particles2}.}
\label{tab:particles3}
\end{table}%

In this scenario, as the quark and lepton masses are generated from the same soft breaking terms, a correlation between both sectors may be found for simplified cases. The phenomenology of vector-like quarks apply to this model, giving a rough conservative limit of $\mathcal{O}(1)$ TeV to their masses \cite{Aaboud:2017zfn,Aaboud:2018saj}. Furthermore, limits on the scalar leptoquark masses can be found in Refs. \cite{Dorsner:2016wpm,Diaz:2017lit,Schmaltz:2018nls,Vignaroli:2018lpq}.

\subsection{Option 3: All vector-like fields are color singlets}

We finally let us analyze the possibility that none of the vector-like fields carry color charge. In this case specified in Table~\ref{tab:particles4}, the same fermions that mediate the lepton masses can also enter the quark masses, i.e. $F_i \equiv \Psi_i$. Leptoquark scalars are  also needed. The scalar sector can be simplified by setting $\Delta = \Phi_2$ and  $\Phi_6 = \Phi_5$. Likewise the previous case, correlations between the lepton and quark masses can sometimes be found. For the limits on the vector-like lepton masses we refer to \cite{Falkowski:2013jya,Kumar:2015tna}.

\begin{table}[h!]
\begin{center}
\begin{tabular}{| c || c | c | c | c || c |}
  \hline
  &  \hspace{0.1cm} Fields \hspace{0.1cm}  &  \hspace{0.4cm} $SU(3)_C \times SU(2)_L \times U(1)_Y$ \hspace{0.4cm}  &  \hspace{0.4cm} $Z_4$ \hspace{0.4cm}  &  \hspace{0.4cm} $U(1)_{X}$ \hspace{0.4cm}  &  \hspace{0.4cm} Comments \hspace{0.4cm}  \\
\hline \hline
\multirow{3}{*}{ \begin{turn}{90} Fermions \end{turn} }
    &  ($F_{1L}$, $F_{1R}$) &  ($\mathbf{1}$, $\mathbf{1}$,    0) &  ($1$, $-1$) &     $x$ & Vector-like $\nu_R$ \\
    &  ($F_{2L}$, $F_{2R}$) &  ($\mathbf{1}$, $\mathbf{1}$,    1) &  ($1$, $-1$) &    $-x$ & Vector-like $e_R$ \\
    &  ($F_{3L}$, $F_{3R}$) &  ($\mathbf{1}$, $\mathbf{2}$, -1/2) &  ($1$, $-i$) &  $-x/2$ & Vector-like doublet $L$ \\ 
\hline \hline
\multirow{12}{*}{ \begin{turn}{90} Scalars \end{turn} } 
    &  $S_1$ &            ($\mathbf{3}$, $\mathbf{1}$, -1/3) &  $-1$ &    $-x$ & Scalar $d$    \\
    &  $S_2$ & ($\mathbf{\overline{3}}$, $\mathbf{2}$, -1/6) &  $-1$ &     $x$ & Scalar $Q$    \\
    &  $S_3$ &            ($\mathbf{1}$, $\mathbf{2}$,  1/2) &   $i$ &  $3x/2$ & Inert doublet \\
    &  $S_4$ & ($\mathbf{\overline{3}}$, $\mathbf{2}$, -1/6) &   $i$ &  $-x/2$ & Scalar $Q$    \\
    &  $S_5$ &            ($\mathbf{1}$, $\mathbf{2}$,  1/2) &  $-i$ &   $x/2$ & Inert doublet \\
    &  $S_6$ &            ($\mathbf{3}$, $\mathbf{2}$, -5/6) &  $-i$ &   $x/2$ & - \\
    &  $\Xi$ &            ($\mathbf{3}$, $\mathbf{2}$, -5/6) &  $-1$ &     $x$ & - \\
    \cline{2-6}
    &  $\Phi_1$ & ($\mathbf{1}$, $\mathbf{1}$,   -1) &  $-1$ &    $-x$ & Scalar charged singlet \\
    &  $\Phi_2$ & ($\mathbf{1}$, $\mathbf{2}$,  1/2) &  $-1$ &     $x$ & Inert doublet \\
    &  $\Phi_3$ & ($\mathbf{1}$, $\mathbf{2}$,  1/2) &   $i$ &  $3x/2$ & Inert doublet \\
    &  $\Phi_4$ & ($\mathbf{1}$, $\mathbf{2}$,  1/2) &   $i$ &  $-x/2$ & Inert doublet \\
    &  $\Phi_5$ & ($\mathbf{1}$, $\mathbf{2}$,  1/2) &  $-i$ &   $x/2$ & Inert doublet \\
    \hline
  \end{tabular}
\end{center}
\caption{Charge assignments for the fermions and scalars that participate in the diagrams \mbox{Figs.~\ref{fig:quarksD}-\ref{fig:nudirac}} considering no vector-like colored fermions. As before, new colored scalars with $U(1)_{X}$ charges appear. Other notations are the same as in Table~\ref{tab:particles2}.}
 \label{tab:particles4}
\end{table}%

All the examples of models discussed here represent just a small set of the possibilities. We have restricted ourselves to the models containing fields up to the $SU(3)_C$-triplets and the $SU(2)_L$-doublets with hypercharge below two. Other particle contents with both colored and non-colored vector-like fermions can be of interest too, as well as more exotic fields, for which a rich phenomenology may exist. It is worth mentioning that a scalar leptoquark with the SM quantum numbers ($\mathbf{\overline{3}}$, $\mathbf{1}$, -1/3) can give rise to proton decay at tree level \cite{Arnold:2012sd, Assad:2017iib}. Usually, proton decay is forbidden by imposing either a discrete cyclic symmetry \cite{Arnold:2013cva}, or a $U(1)$ symmetry, which can be either local \cite{Assad:2017iib} or global. As discussed in the next section, a global $U(1)_{X}$ symmetry is implemented in our framework. Under this symmetry the SM fermionic fields are neutral, whereas the scalar leptoquarks are charged. As a result, the tree-level operators that induce proton decay are forbidden.
\\

Moreover, concerning the particle content, the additional fields modify, in principle, the running of the corresponding gauge couplings. For a gauge group $G=SU(N)$, the running of the coupling constant $g$ is given by:
\begin{equation} \label{eq:beta}
    16 \pi^2 \frac{1}{\mu} \frac{dg}{d\mu} = \beta g^3, \quad \text{with} \quad \beta = \frac{1}{3} \left( \sum\limits_{\#\text{dof}} T(R) - 11 C_2(G) \right) - ... \; ,
\end{equation}
where the sum is done over the degrees of freedom (dof),\footnote{Note that Dirac fermions have four degrees of freedom.} $T(R)$ is the Dynkin index of the representation of the corresponding dof and $C_2(G)=N$, the quadratic Casimir of $G$.\footnote{For $U(1)$, $T(R)$ should be substituted by the squared of the charge and $C_2(G)$ set to zero.} The ellipsis $(...)$ accounts for two-loop corrections induced by the Yukawa couplings, still proportional to $g^3$, but suppressed by an extra factor $1/16\pi^2$ \cite{Popov:2016fzr}.

Going back to our example models, since we are only considering representations up to doublets with small hypercharges, the runnings for $SU(2)_L$ and $U(1)_Y$ are not an issue and yield Landau poles beyond $10^{16}$ GeV in the worst case scenario, while for $SU(3)_C$ there are not enough new color degrees of freedom in the models to drive the corresponding value of $\beta$ positive. From \eqref{eq:beta} it follows that in order to have a Landau pole below the Planck scale, i.e. $\Lambda_{LP} \lesssim 10^{18}$ GeV, $\beta$ should be larger than roughly $2$, for which at least $26$ ($53$) new color triplet Weyl fermions (scalars) are approximately needed.

To close this section, we proceed to comment on signals that can be use to distinguish, from an experimental point of view, the example models we described. Regarding models 1 and 2, one can consider the production of a pair of vector-like fermions in association with the SM like Higgs via vector boson fusion. Given that in models 1 and 2 such vector-like fermions are colored whereas in the model 3, they do not, the process $pp\to h F_i F_i$ ($i=1,2,3$) can be used to differentiate them. Besides that, given that such vector-like fermions are pair-produced and considering that they decay into a SM charged fermion and a heavy scalar (which in several cases is electrically neutral), one has that the observation of an excess of events with respect to the SM background in the dijet final state at the LHC, can be a signal of support of models 1 and 2, whereas the observation of an excess of events in opposite sign dileptons can be a collider signature of model 3. Furthermore, from a dark matter candidate point of view, it is worth mentioning that models 1 and 3 have fermionic dark matter candidates, whereas model 2 does not.

\section{About the soft breaking} \label{sec:softbreaking}

In this section we explain the structure of the soft breaking terms, motivating the introduction of the global $U(1)_{X}$ symmetry in our models. Given a set of symmetries and fields, in order to be sure that the theory is renormalizable, one should include all the terms allowed by the symmetry with mass dimensions up to four. When some symmetry is softly broken, i.e. by introducing explicit breaking terms with dimension-3 or below (soft terms) in the interaction Lagrangian, no divergences appear for dimension-4 interaction terms (hard terms). However, divergences may arise, if one does not add all the soft terms that break the symmetry in the same way, therefore, spoiling renormalizability. A conservative approach, to be sure that the theory is still renormalizable after soft-breaking, is to include in the Lagrangian every soft term allowed by the remnant symmetries.

\begin{figure}[h!]
    \centering
    \includegraphics[scale=0.9]{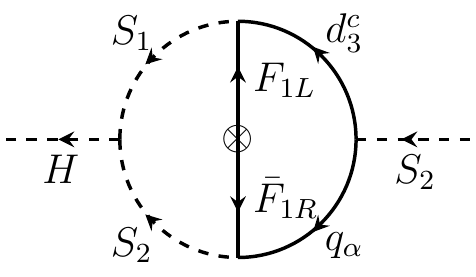}
    \caption{Two-loop diagram that generates an infinite contribution to the soft term $H S_{2}$. Note that this diagram is built directly from the one-loop bottom quark mass \fig{fig:quarksD}~(left) by closing the fermion lines. This diagram implies that once the soft term $\bar F_{1R} F_{1L}$ ($\otimes$) is added, $H S_{2}$ should be included too. If not, the resulting theory would not be renormalizable.}
    \label{fig:sbt}
\end{figure}

In all the radiative models described above, the $Z_{4}$ symmetry is broken by adding soft terms depicted as $\otimes$ in the diagrams. In principle, as the symmetry is broken, we have to write every soft term invariant under the SM. Considering, for example, the one-loop diagram describing the bottom quark mass \fig{fig:quarksD}~(left) with the field content of Table~\ref{tab:particles2}, $Z_{4}$ is broken to a $Z_{2}$ by the soft term $\bar F_{1R} F_{1L}$. This implies that the soft term $H S_{2}$ has also to be included in the Lagrangian in order to absorb infinite contributions arising from diagrams like \fig{fig:sbt}. 

Therefore, $S_2$ will mix with the SM Higgs boson $H$ and generate a tree-level mass for quarks through the term $\bar F_{1R} q_\alpha S_2$. We add the $U(1)_{X}$ symmetry to protect the models from these issues, requiring that it forbids the mixing of the internal loop particles with the SM ones. The condition that the vertices given in Figs.~\ref{fig:quarksU}-\ref{fig:nudirac} are allowed, while the undesired soft terms are not, defines a system of linear equations for the $U(1)_{X}$ charges. Its solution, considering that the SM fields are not charged, is given in Tables~\ref{tab:particles2}-\ref{tab:particles4} in terms of a free parameter $x$, with $x\neq 0$.

It is worth mentioning that the existence of soft terms as $H S_{2}$ is completely model dependent. In our case, it is due to our choice of the quantum numbers of $S_2$ in section~\ref{sec:particles 1}, which transforms as the Higgs under the SM. For other choice of the quantum numbers of the internal fields, such couplings may not exist and one can decouple the internal particles from the SM without, in principle, any new symmetry. Nevertheless, the same $U(1)_{X}$ global symmetry we explicitly included here will still be a symmetry of the theory but at the accidental level.

\section{Discussion} \label{sec:discussion}

We proposed a systematic method of construction of renormalizable models of the SM fermion masses with the observed hierarchy originating from the sequential loop suppression mechanism resorting to a single softly broken discrete symmetry. We applied this method to an example model based on $Z_{4}$-symmetry. In this model, non-SM fields and the softly broken $Z_4$ symmetry are devised to generate tree-level top quark mass; 1-loop bottom, charm, tau and muon masses; 2-loop masses for the light up, down and strange quarks as well as for the electron; and effective 4-loop masses for the light active neutrinos. The 4-loop level masses for the light active neutrinos arise from the radiative Type-I seesaw mechanism, where the Dirac mass terms are generated at two loop level. To illustrate the sequential loop suppression mechanism we considered three scenarios: 1) With no colored scalars. 2) With all vector-like fields colored. 3) With no colored vector-like fields. We also discussed self-consistent inclusion of the soft breaking terms needed for the implementation of the proposed mechanism of the sequential loop suppression in the way that undesirable radiatively induced soft terms are suppressed by an accidental $U(1)_{X}$ symmetry.

The mechanism described in section~\ref{sec:model} is general and can be, in principle, apply to abelian and non-abelian discrete symmetry. In this work, we focus on an example with a $Z_4$ symmetry for simplicity, to show how the mechanism works. The same discussion about the generation of SM fermion masses can be extended to non-abelian discrete symmetries and use the same symmetry to motivate the existence of three generations. For this purpose, one should provide a symmetry with a proper breaking chain or having the $Z_N$ as a subgroup of it, for example, the group $\Delta( 3N^2 )$. Nevertheless, a detailed study of this possibility is beyond the scope of this paper and it is left for future works.

Due to this accidental symmetry, the models considered in this paper have several stable scalar and fermionic dark matter candidates. Considering a scenario with a scalar DM candidate, one has to ensure its stability, which can be done by assuming that it is the lightest among the inert scalar particles and is lighter than the exotic fermions. That scalar DM candidate  annihilates mainly into $WW$, $ZZ$, $t\overline{t}$, $b\overline{b}$ and $hh$ via a Higgs portal scalar interaction $\lambda_{DM}\left(H^{\dagger}H\right)\Phi^{\dagger}_{DM}\Phi_{DM}$, where $H$ is the SM Higgs doublet and $\Phi_{DM}$ the scalar DM candidate in our model. These annihilation channels will contribute to the DM relic density, which can be accommodated for the appropriate values of the scalar DM mass and of the quartic scalar coupling  $\lambda_{DM}$, similarly to Refs. \cite{CarcamoHernandez:2016pdu, Bernal:2017xat, CarcamoHernandez:2017kra, Long:2018dun}. Thus, for the DM direct detection prospects, the scalar DM candidate would scatter off a nuclear target in a detector via Higgs boson exchange in the $t$-channel, giving rise to a constraint on the coupling $\lambda_{DM}$. Given the large number of parameters in the scalar potential of the models considered in this work, there is a lot of parametric freedom that allows us to successfully reproduce the current experimental value of the DM relic density. The parameter space of our model consistent with DM constraints will be similar to the one in 
Refs.~\cite{CarcamoHernandez:2016pdu, Bernal:2017xat, CarcamoHernandez:2017kra, Long:2018dun}. 

Finally, it is worth mentioning that the charged exotic fermions in our model can decay into the SM charged fermion and a heavy scalar, which in several cases can be electrically neutral. For that scenario, the heavy electrically neutral scalar can be identified as missing energy. In addition, the heavy charged exotic fermions can be pair-produced at the LHC via Drell-Yan annihilation only, if they are leptons, and via gluon fusion and Drell-Yan mechanism, if they are quarks. Consequently, the observation of an excess of events with respect to the SM background in the dijet and opposite sign dileptons final states at the LHC, can be treated as a signal supporting the models considered in this work. The precise signature of the decays of the exotic fermions depends on details of the spectrum and other parameters of the models considered here. It is worth mentioning that models with vector-like fermions are being tested at the LHC, and the ATLAS collaboration has reported several analyses in this direction, in particular~\cite{Aaboud:2017zfn, Aaboud:2018saj, Aaboud:2018pii}. Currently, the bounds $M_B > 1.22$ TeV and $M_T > 1.31$ TeV for the exotic isosinglet quark masses have been obtained and presented in Ref.~\cite{Aaboud:2018pii}, under the assumption that the exotic quarks can decay to the  SM particles. Specifically, in those studies it is assumed that the vector-like quarks would be first produced at collider experiments via pair-production, a process dominated by gluon fusion mechanism, $gg\rightarrow \bar{B} B \, (\bar{T} T)$. Then, each exotic quark would decay to: $T\rightarrow Wb, Zt, H t$ or $B\rightarrow Wt, Zb, Hb$, where only the third fermion family  has been considered.

A detailed study of the collider phenomenology of the proposed models is beyond the scope of this paper and it is left for future studies.

%%%%%%%%%%%%%%%%%%%%%%%%
\centerline{\bf Acknowledgements}

\medskip
 We are grateful to Martin Hirsch for helpful discussions. C.A., A.E.C.H, S.K. and I.S. are supported by CONICYT-Chile FONDECYT 11180722, CONICYT-Chile FONDECYT 1170803, CONICYT-Chile FONDECYT 1190845, CONICYT-Chile FONDECYT 1180232, CONICYT-Chile FONDECYT 3150472 and ANID PIA/APOYO AFB180002. R.C. acknowledges funding by Spanish grants FPA2017-90566-REDC (Red Consolider MultiDark), FPA2017-85216-P (AEI/FEDER, UE), PROMETEO/2018/165 (Generalitat Valenciana), FPU15/03158 and Beca Santander Iberoam\'erica 2018/19. R.C. would like to thank the USM Department of Physics and CCTVal for their hospitality.

\bigskip

\appendix

\section{Other realizations} \label{sec:app}

Here, 
%In this appendix 
we briefly discuss other possible particle contents apart from those given in section~\ref{sec:particles}. The main purpose is to show different realizations where $U(1)_{X}$ is  not explicitly imposed on theory. 
%Such examples of particle contents we have found involved either higher representations and charges under the Standard Model, or a larger discrete symmetry from the starting point.

As discussed in the main text, the $U(1)_{X}$ 
%explicit 
symmetry is model-dependent and is not related to the loop suppression mechanism described in section~\ref{sec:model}. It plays an auxiliary role helping us to forbid the mixing of the internal loop fields with the SM ones.

In the three options for the realizations of our scenario considered previously, we required that 
%this symmetry 
%because we wanted 
the fields belong to the SM-group representations up to $SU(3)_{c}$-triplets, up to $SU(2)_L$-doublets
%representations up to doublets 
 and have small hypercharges.  After breaking $Z_4$, they mix with the SM fields.
 However, if we relax these limitations on the field assignment and introduce the representations of higher dimensions and larger hypercharges, we  are able to decouple the exotic fields from the SM ones without any extra symmetry like $U(1)_{X}$.
% one relaxes these constraints about the possible quantum numbers,by raising the representations and/or hypercharges of the new fields, these will decouple from the SM without any extra symmetry. 
Nevertheless, 
% However, 
$U(1)_X$ (or its subgroup) will appear as an accidental symmetry of the theory.

In table~\ref{tab:particles5} we show one example of a particle content, which does not require an 
%explicit 
extra symmetry. Compared to the options shown in tables~\ref{tab:particles2}-\ref{tab:particles4}, 

%{\color{cyan}we have that} 
in this case there are no Higgs-like doublets that can 
%get induced 
develop vevs after the $Z_4$ soft-breaking, generating a lower order mass term. 
%The only problematic field may be 
Nevertheless, there is $S_3 \equiv (\mathbf{1}, \mathbf{1}, 0)$ field, which,
%{\color{cyan}since due to its} %that, 
given 
its  quantum numbers, 
%{\color{cyan}it} can have an induced 
can acquire vev via a one-loop coupling with $H^\dagger H$. However, $S_3$ couples only to the BSM fermions $F_{1,2}$ and $\Psi_{1,2}$ and, therefore, cannot generate by itself a mass term for the Standard Model fermions. It would just contribute to a diagram similar to Figs.~\ref{fig:quarksD} and \ref{fig:chargedleptons},  where $S_3$ develops a one-loop induced vev. In this case, the loop order would be the same, but it would be suppressed by the smallness of the induced vev. Consequently, this contribution would be a correction to those, which are 
%{\color{cyan}ones} 
generated by the diagrams in Figs.~\ref{fig:quarksD} and \ref{fig:chargedleptons}. Note that there is still an accidental $U(1)$ symmetry (or a subgroup of it), as explained before. This symmetry is related to the fact that after breaking the $Z_4$, the fields inside the loop, except $S_3$, are coupled to the SM only via the vertices depicted in the mass diagrams, i.e. they cannot  decay solely to SM particles.

\begin{table}[h!]
\begin{center}
\begin{tabular}{| c || c | c | c | c || c |}
  \hline
  &  \hspace{0.1cm} Fields \hspace{0.1cm}  &  \hspace{0.4cm} $SU(3)_C \times SU(2)_L \times U(1)_Y$ \hspace{0.4cm}  &  \hspace{0.4cm} $Z_4$ \hspace{0.4cm} \\
\hline \hline
\multirow{6}{*}{ \begin{turn}{90} Fermions \end{turn} }
    &    ($F_{1L}$,    $F_{1R}$) &  ($\mathbf{3}$, $\mathbf{1}$, -4/3) &  ($1$, $-1$) \\
    &    ($F_{2L}$,    $F_{2R}$) &  ($\mathbf{3}$, $\mathbf{1}$,  5/3) &  ($1$, $-1$) \\
    &    ($F_{3L}$,    $F_{3R}$) &  ($\mathbf{3}$, $\mathbf{1}$,  5/3) &  ($1$, $-i$) \\ 
    & ($\Psi_{1L}$, $\Psi_{1R}$) &  ($\mathbf{1}$, $\mathbf{1}$,   -2) &  ($1$, $-1$) \\
    & ($\Psi_{2L}$, $\Psi_{2R}$) &  ($\mathbf{1}$, $\mathbf{1}$,    1) &  ($1$, $-1$) \\
    & ($\Psi_{3L}$, $\Psi_{3R}$) &  ($\mathbf{1}$, $\mathbf{1}$,    0) &  ($1$, $-i$) \\
\hline \hline
\multirow{5}{*}{ \begin{turn}{90} Scalars \end{turn} } 
    & $S_1$ & ($\mathbf{1}$, $\mathbf{1}$,    1) &  $-1$ \\
    & $S_2$ & ($\mathbf{1}$, $\mathbf{2}$, -3/2) &  $-1$ \\
    & $S_3$ & ($\mathbf{1}$, $\mathbf{1}$,    0) &   $i$ \\
    & $S_4$ & ($\mathbf{1}$, $\mathbf{1}$,   -1) &   $i$ \\
    & $S_5$ & ($\mathbf{1}$, $\mathbf{1}$,   -3) &  $-i$ \\
    & $S_6$ & ($\mathbf{1}$, $\mathbf{1}$,    2) &  $-i$ \\
    \hline
  \end{tabular}
\end{center}
\caption{Charge assignments for the fermions and scalars contributing to the diagrams Figs.~\ref{fig:quarksD}-\ref{fig:nudirac}. In the brackets of the $Z_{4}$-column we indicate the corresponding charges of the left-handed and right-handed vector-like fermions, respectively. Here $S_i \equiv \Phi_i$ and $\Xi=\Delta=S_2$.}
 \label{tab:particles5}
\end{table}%

It is worth mentioning that there 
%other particle contents 
exist other particle contents apart from table~\ref{tab:particles5}, which allow 
%can 
avoiding the model-dependent $U(1)_X$ symmetry. Nevertheless, a naive choice of the SM charges may lead to either a Landau pole, or to a stable charged and/or colored particle, with catastrophic phenomenological implications.\footnote{If all the fields in the loop are decoupled from the SM, the lightest field would be stable. Then, if none of the multiplets has a electrically neutral colorless component, the model would lead to a stable charged and/or colored particle which is forbid by cosmology \cite{Langacker:2011db}.}
\\

We would like to close this appendix by pointing out another way to avoid the $U(1)_X$ symmetry. Note that the full $U(1)$ group is actually not needed to decouple the fields running in the loop from the Standard Model. A subgroup of $U(1)$ may do the job, i.e. a discrete $Z_n$ should be enough. This is similar to the well-known neutrino mass models, like the \textit{scotogenic model} \cite{Ma:2006km}. 

In sections~\ref{sec:quarks} and \ref{sec:leptons}, we focused on $Z_4$ as an example to generate the SM fermion masses, because it is the smallest $Z_n$ symmetry, which can work with the mechanism described in the present paper. However, this implied that, in order to realize the model with a particle content with small hypercharges and representations, one should explicitly add a $U(1)$ symmetry, which otherwise would have been accidental. For example, starting with a $Z_{4n}$ discrete symmetry with $n>1$, would lead to the same set of diagrams by softly breaking only a $Z_4$ subgroup of $Z_{4n}$. Then, a remnant $Z_n$ symmetry survives, which can be used to forbid the mixing of the internal fields with the SM, similarly to $U(1)_X$. For instance, if we allow $S_{3,5}$ ($\Phi_{3.5}$) to decay, $Z_8 \supset Z_{4}$ should be enough, so a $Z_2$ residual symmetry survives ensuring the hierarchy for an arbitrary particle content. While, if we want everything in the loop to be stable, it can be done with a remnant $Z_3$ discrete symmetry, starting for example from a $Z_{12} \supset Z_{4}$.

% Bibliography
\bibliographystyle{t1}
\bibliography{MyBibTexDatabase}

\end{document}